\documentclass[10pt]{article}
\usepackage{arxiv}

\usepackage[letterpaper]{geometry}
\usepackage[]{xcolor}
\usepackage[fleqn,intlimits]{amsmath}
\usepackage{graphicx}
\usepackage{}
\usepackage{wrapfig}

\usepackage[ruled,vlined]{algorithm2e}
\usepackage{xcolor}

\usepackage[utf8]{inputenc} 
\usepackage[T1]{fontenc}    
\usepackage{hyperref}       
\usepackage{url}            
\usepackage{booktabs}       
\usepackage{amsfonts}       
\usepackage{nicefrac}       
\usepackage{microtype}      
\usepackage{cleveref}       
\usepackage{graphicx}
\usepackage{doi}

\SetCommentSty{mycommfont}

\SetKwInput{KwInput}{Input}                
\SetKwInput{KwOutput}{Output}              

\usepackage{epstopdf}
\usepackage{tikz}
\usepackage{hyphenat}

\def\Ham{\mathcal H}
\def\path{\mathcal P}
\def\wij{\omega_{ij}}
\def\C{\mathcal C}
\def\N{\mathcal N}
\def\D{\mathcal D}

\title{City path tomography: reconstructing square road network from artificial users mobile phone data} 

\newcommand\shorttitle{City path tomography: network reconstruction from mobile phone data} 
\author{Andy Rodríguez Lorenzo \\
		\texttt{androdn96@gmail.com} \\
    \AND
	Alejandro Lage-Castellanos
	\thanks{Corresponding author.} \\
		\texttt{ale.lage@gmail.com} \\
		\\
	Group of Complex Systems, Physics Faculty, University of Havana, Cuba }

\hypersetup{
pdftitle={City path tomography},
pdfsubject={cond-mat.stat-mech},
pdfauthor={Andy Rodriguez Lorenzo, Alejandro Lage-Castellanos},
pdfkeywords={Statistical physics, Inference, Inverse models,
Computer modeling and simulation, Telecommunications, Transport network},
}

\begin{document}

\maketitle

\begin{abstract}
Population mobility can be studied readily and cheaply using cellphone data, since people's mobility can be approximately mapped into tower-mobile registries. 
We model people moving in a grid-like city, where edges of the grid are weighted and paths are chosen according to overall weights between origin and destination. Cellphone users leave sparse signals in random nodes of the grid as they move by, mimicking the type of data collected from the tower-cellphone interactions. From this noisy data we seek to build a model of the city, {\it i.e.} to predict probabilities of paths from origin to destination. We focus on the simplest case where users move along shortest paths (no loops, no going backwards). In this simplified setting, we are able to infer the underlying weights of the edges (akin to road transitability) with an inverse statistical mechanic model.   
\end{abstract}

%
%
%
%
%
%
%

\section{Introduction}
\label{sec:intro}

\nohyphens{Population size, concentration and mobility have been constantly growing during most of human history, and specially so in the last 200 years. These changes have impacted geography, demography and social dynamics in all too many ways, sometimes at a pace faster than the adaptability of the systems. 
Understanding human mobility patterns at city, country, and international scales is useful for decision-making related with urban planning, transport and dealing with infectious diseases \cite{asgari_survey_2013}.}

Many different sources have been used to obtain information about the population mobility, a considerable part of those are expensive or slow.
It is possible to find human mobility investigations carried out with the information of a national census \cite{becker_human_2013} or bank notes \cite{brockmann_scaling_2006}. 
Also GPS devices have been used \cite{gong_gpsgis_2012}. 
GPS data provides accurate measures of position in outdoors with the drawback of less accuracy because of weak signals in indoor scenarios and the tendency of users to turn off the service when isn't needed due to battery consumption. 

The use of telecommunication data allows researchers to perform these studies in a faster and cheaper way \cite{huang_transport_2019}.
Large telecommunications companies, private applications, and network providers collect and store enormous quantities of data on users of their products and services \cite{toole_path_2015}. 
A massive number of phone calls can be processed and arranged in records that might include information of the user (usually anonymized), the position of the device or at least the radio base to what the user's device is connected and a time stamp \cite{chen_promises_2016}.

The use of telecommunication data opens a new era in the human mobility studies \cite{asgari_survey_2013}.
This kind of data is now available to researchers in many countries, even in developing countries \cite{calabrese_urban_2014}.
Due to its vast amount it is possible to improve the results in comparison with traditional sources as surveys and census and start new trends in human mobility fields. 

However, a key aspect hindering its applicability is that of privacy\cite{shin_privacy_2012}, resulting on a limited access to raw data (even after anonymization) for researchers and foremost for end-users as decision-makers. An alternative approach is that of building models of social behavior from this data, such that the models can be shared without privacy concerns.


\nohyphens{In this paper we formalize the problem of understanding city connections (road usability or transitability) from analyzing noisy and sparse cellphone data in an artificial city. We model the city as a weighted 2D grid describing the connections among parts of the city. Some papers already have taken graph based approaches. For instance in \cite{tesselkin_estimation_2017,pourmoradnasseri_od-matrix_2019} a transportation network graph is presented to calculate origin destination matrices using Markov models. Instead, we map our reconstruction problem, that we call {\bf city path tomography}, as an inverse statistical mechanics model. We develop a procedure to infer weights for edges of the city from the sparse cellphone data using a gradient descent algorithm. The main limitation of the actual proposal is related to the capability of fast computation of the partition functions that we are able to solve for 2D-grids, but remain a complicated issue for more heteromorphic cities.}

Origin-destination (O-D) matrices are the standard information taken from mobility studies. Not withstanding its relevance, it provides no clue on how people move between the given O-D points. City path tomography is somehow a complementary approach. It aims to produce a probabilistic model of the usage of different paths between given O-D points in a city. 



\nohyphens{The paper is organized as follows: section PATH TOMOGRAPHY describe the city and its connection to the artificial paths of users. In section MAX-LIKELIHOOD we discuss an efficient procedure to infer the weights of the city based on the incomplete path data. In section RESULTS we apply this method to some artificial datasets and discuss its performance. Finally CONCLUSIONS are drawn.}
\section{Path tomography}
\label{sec:ptom}

We consider a simplified version of city as a squared lattice graph $\C =~(V,E,W)$, a 2D grid of weighted nodes and/or edges. 
Let $L$ be the side of the square city, we label the nodes in the city as $V={1,\ldots,L^2}$, and define paths, from start $h_0$ to end $h_n$ as a sequence of contiguous edges in the graph:
\begin{equation}
	\path = \{ h_0,h_1,\ldots,h_{n-1},h_n\} \:\: \mbox{where }\forall_i (h_i,h_{i+1}) \in E(\C). \label{eq:path}
\end{equation}
\noindent Furthermore, we will only consider paths that take the traveler closer to destination at every step, for example if $h_n$ is above/(to the right) of node $h_0$, then $h_{i+1}$ is always above or to the right of node $h_i$ in the sequence. 
This also means that if $h_0$ and $h_n$ are $p$ step distant in the y-axis and $q$ steps in the x-axis, then the total number of nodes visited in every non returning path is $n=p+q$, and the amount of such paths is equal to:
\begin{equation}
	\N(h_0,h_n) = \binom{n}{p} \label{eq:npaths}
\end{equation}

\begin{figure}
	\centering
	\begin{tikzpicture}[scale=0.8]
	\foreach \x in {0,...,4}
	\foreach \y in {0,...,4}
	{
		\fill (\x,\y) circle (3pt);
	}
	\draw[>=latex ,<->] (0,-0.5) -- (4,-0.5);
	\node at (2,-1) (n1) {L};
	\draw[>=latex ,<->] (-0.5,0) -- (-0.5,4);
	\node at (-1,2) (n2) {L};
	\draw[>=latex,color=red ] (1.1,3) -- (1.9,3);
	\node at (1,2.6) (ai) {$a_i$};
	\node at (2,2.6) (aj) {$a_j$};
	\node at (1.5,3.3) (w) {$\omega_{ij}$};
	\end{tikzpicture}
	\label{city}
	\caption{Example of 2D city grid, with nodes and links weighted}
\end{figure}
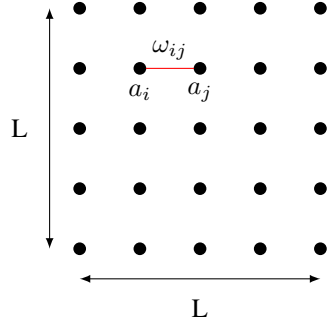

In figure~\ref{city} is shown an example of this kind of 2D grid using $L=5$.

In order to build a statistical model for the trajectories of users in the city, we define the Hamiltonian (cost function) as the total weight of a path as follows:
\begin{equation}
	\Ham(\path) = \sum_{i\in \path \setminus \{h_0 \bigcup h_n\}} a_i + \sum_{(ij)\in \path} \wij \qquad  \label{eq:ham}
\end{equation} 
\nohyphens{where the first sum runs over nodes weights while the second over the links weights.}

Be noticed that the first and last node weight wasn't added in the Hamiltonian. 
The probability of a given path between points $h$ and $m$ in the grid is therefore given by
\begin{equation}
	P(\path) = \frac 1 {Z_{hm} } \exp{\left(-\beta \Ham(\path)\right)} \label{eq:PP}
\end{equation}
where the normalization $Z_{hm}$ ensures that, constrained to the same origin $h$ and destination $m$, the set of all non returning paths $\path(h,m)$ has probability 1:
\begin{equation}
\sum_{\path\in\path(h,m)} P(\path)= 1 \Rightarrow Z_{hm}  = \sum_{\path\in\path(h,m)} \! \exp{\left(-\beta \Ham(\path)\right)}. \label{eq:Z}
\end{equation}

\subsection{Telecom data}

\nohyphens{For the sake of our model, assume that towers exist at every node of our grid-like city, and that travelers with a mobile phone can leave registries of phone-tower communication in the nodes along their path. However, registry data from these paths is usually incomplete, and they usually consist on only but a few of the nodes visited. For instance, paths could look like}

\begin{eqnarray*}
\mathcal{D} = \left\{\begin{array}{l}
\left[95, 46, 48, 14, 42, 128, 116, 117\right] \\
\left[14, 61, 68\right] \\
\left[95, 128, 112, 117 \right] \\
\ldots
                     \end{array} \right. 
\end{eqnarray*}
where each row correspond to a single user, and numbers correspond to tower ID's. As you can see from the first and third lines, the amount of signals left from a trip could vary widely, even between same origin and destination. We break down every multi-tower path to its minimum informative value, which is a set of triplets
\begin{eqnarray*}
\left.\begin{array}{c}
\left[95, 46, 117\right] \\
\left[95, 48, 117\right] \\
\left[95, 14, 117\right] \\
\left[95, 42, 117\right] \\
\left[95, 128,117\right] \\
\left[95, 116, 117\right] 
\end{array} \right\} & 1/6& \\
\left.\left[14, 61, 68\right]\:\:\: \right\} \: & 1& \\
\left.\begin{array}{c}
\left[95, 128, 117\right] \\
\left[95, 112, 117\right] 
\end{array}\right\} & 1/2& 
\end{eqnarray*}
When a given path has more than one (say $k$) intermediate points, each of its triplets is weighted as $1/k$ to avoid an statistical bias on the path due to its length.

We consider the travelers decisions around the city to be given by the model \eqref{eq:PP}, and we assume that given a path as \eqref{eq:path}, every visited node leaves a signal in the mobile-phone records with a probability $\eta <1$.

\nohyphens{The path tomography problem (PTP) that we are interested in is the following:}

{\bf Definition [PTP]: }given a large set of incomplete paths $\mathcal{D}$ of users  in the city, infer the set of weights in $W$ that define model \eqref{eq:PP}.



Although we can consider node-weighted or edge-weighted cities, or even both, it can be shown that the probabilistic model given by \eqref{eq:PP} is invariant to local transformations of the type
\begin{equation}
a_i \to a_i +\Delta \qquad \forall_{j\in \partial j}\:\: \wij \to \wij -\frac \Delta 2.
\end{equation}
This implies that a model with edge-weights is equivalent to the more general model with both edge and node weights, since every node weight can be freely set to zero by taking $\Delta_i = -a_i$. However, we will keep the model in full generality.

We will proceed with the node-weight graph first, and extensions to edge-weighted graphs is immediate. Given that someone is known to have moved from (non-adjacent) nodes $i$ to $j$, the probability that it has passed through node $k$ is given by:
\begin{equation}
	P(k|i,j) = \frac {Z_{ik} \: e^{-\beta a_k}\: Z_{kj}}{ Z_{ij}}. \label{eq:Pikj}
\end{equation}

\section{Max-likelihood inference}

\nohyphens{The model \eqref{eq:PP} considers probabilities of different paths between a given origin $O$ and destination, but do not take into account (nor care) the frequency of each $O-D$ pairs. Therefore we build a likelihood not in terms of full probabilities $p(i,k,j) = P(k|i,j) P(O=i,D=j)$, but rather in terms of the conditionals \eqref{eq:Pikj}:}
\begin{align}
	 L(\D) &= \sum_{(ikj) \in \D} \log P(k|i,j) \label{eq:log_lik} \\
	 &= \sum_{(ikj) \in \D} -\beta a_k + \log Z_{ik} +\log Z_{kj} -\log Z_{ij} \label{eq:log_lik_expanded}
\end{align}
 
If $n_{ij}^O$, $n_{ij}^D$ and $n_{ij}^m$ are the number of elements in $\D$  of the type $(i,j,*)$, $(*,i,j)$ and $(i,*,j)$ respectively, then we define the $n_{ij} = n_{ij}^O +n_{ij}^D -n_{ij}^m$ and
\begin{equation}
	L(\D) = \sum_{(ij) } n_{ij} \log Z_{ij}  - \beta \sum_{k} m_k a_k
	\label{eq:log_lik_reorganized}
\end{equation}
\noindent where all information from the data is subsumed in the coefficients $n_{ij}$. Maximization of the log-likelihood is achieved when its derivatives with respect to the cost function parameters is zero.
Considering that weight $a_k$ appears in the exponential of paths that contain $k$ as an intermediate point, we have:
\begin{equation}
	\frac 1 \beta  \frac{ \partial L(\D)}{\partial a_k}  =  -\sum_{i-j\ni k} n_{ij} \left( \frac 1 Z_{ij} Z_{ik} e^{-\beta a_k} Z_{kj}\right)-m_k \label{eq:grad_nodes}
\end{equation}
The gradient respect to the edges weights $w_{kk'j}$ is:
\begin{align}
	\frac 1 \beta  \frac{ \partial L(\D)}{\partial w_{kk'}}  =  -\!\sum_{i-j\ni (k,k')}  n_{ij}  \left( \frac 1 Z_{ij} Z_{ik} \! e^{-\beta (a_k +w_{kk'} +a_{k'})  }  
	 Z_{k'j}\right)\label{eq:grad_links}
\end{align}

A fast computation of this gradient, in order to implement gradient descent method, requires a clever way to evaluate all the $Z_{ij}$ functions for every pair of possible origin and destination in the city.

\subsection{Gradient fast computation}

As explained in the previous section from a data set and a graph we aim to find the weights that maximize the log-likelihood expressed in \eqref{eq:log_lik}.
Then, the log-likelihood gradient should be computed respect to each type of weight used, let say nodes \eqref{eq:grad_nodes} and links \eqref{eq:grad_links}.

The first step from the gradient computation is to extract the information from the data. This information is condensed in the values of $n_{ij}$ and $m_i$ as explained in the text before \eqref{eq:log_lik_reorganized}. 

The computation of the gradient in terms of a node weight in \eqref{eq:grad_nodes} (link weight in \eqref{eq:grad_links}), requires a sum over all $O-D$ pairs that include that node (link) as a passing point of at least one non returning path from $O$ to $D$. Considering that the city has $n$ nodes, the number of $O-D$ pairs grow as $n^2$, and those sums are not extremely large. The list of $O-D$'s corresponding to every parameter to be extremized, is computed only once at the beginning.


Starting from a random initialization of the weights we apply a gradient descent as is described in algorithm 1.

In the algorithm, these quantities are computed at initialization time before the epoch loop and do not change over the gradient steps:
\begin{itemize}
	\item $\forall_{i,j} n_{ij}$ and $\forall_k m_k$ coefficients computed from data.
	\item possible$\_$nodes$\_$OD: The set of possibles origin-destination for paths including each node.
	\item possible$\_$links$\_$OD: The set of possibles origin-destination for paths including each link.
	\item Initial values of the weights $\omega^0$ and/or $a^0$, chosen randomly.
\end{itemize}
 
 \begin{algorithm}[H]
 	\DontPrintSemicolon
 	{\small Algorithm 1. Gradient descent for city path tomography}
 	\\
 	\KwInput{$n$, $m$, $\omega^0$, $a^0$}
 	\KwOutput{$\omega$, $a$}
 	\ForEach{epoch}{
 		\tcc{Repeat until gradient convergence}
 		Z = Calc\_Part\_Func($\omega^i$, $a^i$) \tcc{$i$: epoch index}
 		\For{k $\mathrm{in}$ nodes}{
 			temp=0
 			
 			\For{i-j $\mathrm{in}$ $\mathrm{possible\_nodes\_OD[\textit{k}]}$}{
 				temp +=  $\dfrac{n_{ij}}{Z_{ij}} Z_{ik} e^{-\beta a_k} Z_{kj}$
 				\tcc{Summation term in equation \eqref{eq:grad_nodes}}
 			}
 			gradient\_$a$[\textit{k}] = -1$\times$temp - m$\mathrm{[\textit{k}]}$	
 		}
 		\tcc{At this point log-likelihood gradient respect to $a$ was computed}
 		
 		\For{$\mathrm{(\textit{k,k'})}$ $\mathrm{in}$ links}{
 			temp=0
 			
 			\For{i-j $\mathrm{in}$ $\mathrm{possible\_links\_OD(\textit{k,k'})}$}{
 				temp += $\dfrac{n_{ij}}{Z_{ij}}Z_{ik} \: e^{-\beta (a_k +w_{kk'} +a_{k'})} \: Z_{k'j}$
 				\tcc{Summation term in equation \eqref{eq:grad_links}}
 			}
 			gradient\_$\omega$[$\mathrm{(\textit{k,k'})}$] = temp	
 		}
 		\tcc{At this point log-likelihood gradient respect to $\omega$ was computed}
 		Update weights:
 		$a^{i+1} = a^{i}+\textrm{learning\_rate}\times \textrm{gradient\_}a$
 		$\omega^{i+1} = \omega^{i}+\textrm{learning\_rate}\times \textrm{gradient\_}\omega$
 		\tcc{Alternatively, an adaptative gradient descent algorithm can be use to compute $a_{i+1}$ and $\omega_{i+1}$}
 	}	
 \end{algorithm}

 However, changing the weights $a$ and $\omega$ do affect the partition functions used in the computation, and it is carried by the function Calc\_Part\_Func($\omega^i$, $a^i$) at the beginning every loop of the gradient descent algorithm. This procedure is described in algorithm 2.

\begin{wrapfigure}{r}{0.4\textwidth} 
	\includegraphics[]{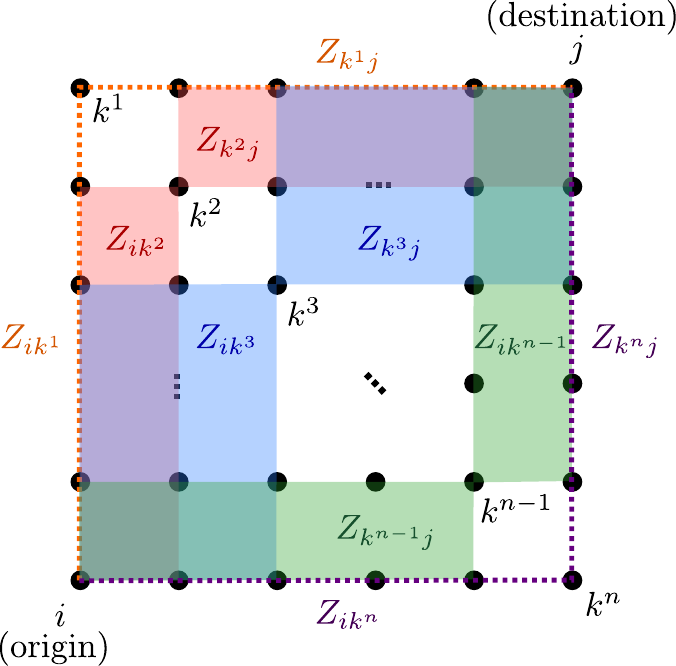}
	\caption{\label{fig:Z_sum} A possible selection of $\mathcal{S}$ and description of partition functions involved}
\vspace{-1cm}
\end{wrapfigure}

Updating the partition functions is the time consuming step in this algorithm. Generally speaking, partition functions are hard, since they typically imply sums over a combinatorial number of states in a model. However, the regular grid-like geometry of the artificial city used and the assumption that only non-returning paths are considered, allow for an efficient computation of the partition function. In particular, the partition function for an origin-destination pair $i - j$ can be split into the partition functions of a decomposition of the paths, $i - k$ and $k - j$ for example. This leads to:
 \begin{equation}
 Z_{ij}=\sum_{k \in \mathcal{S}} Z_{ik}\times \textrm{e}^{-\beta a_k} \times Z_{kj}
 \end{equation}
in this equation $\mathcal{S}$ is a set of points with the same city block distance to $i$ and the same city block distance to $j$. The figure \ref{fig:Z_sum} shows a possible selections of $k$ values to create a set $\mathcal{S}$ and the areas of the paths which each partition function in the summation represent.
 
This results brings two interesting facts. 
Considering the set $\mathcal{S}$ with points only one step away from $j$ and a fixed number of steps away for $i$, any value $Z_{ij}$ can be calculated using:
\begin{equation}
Z_{ij}=Z_{i{j'}}\times \textrm{e}^{-\beta (a_{j'}+\omega_{{j'}j})} + Z_{i{j''}}\times \textrm{e}^{-\beta (a_{j''}+\omega_{{j''}j}))}
\label{eq:rec}
\end{equation}

\noindent where $j'$ and $j''$ are the nodes closer to $j$ in the directions to approach to $i$. With this in mind, recursivity can be used.
Note that in \eqref{eq:rec} $Z_{i{j'}}$ and $Z_{i{j''}}$ are unknown but can be computed with the same idea.
Using this recursive strategy implies that due to the calculation of $Z_{ij}$ are computed partition functions with origin in $i$ and destination each time closer to $i$.  
Then, the calculation of $Z_{ij}$ leads to the computation of every $Z_{i\gamma}$, where $\gamma$ is every node in the rectangle delimited by $i$ and $j$ in the city grid.

\begin{algorithm}[H]
	\DontPrintSemicolon
	{\small Algorithm 2. Procedure to calculate partition functions}
	\\
	\textbf{procedure} Calc\_Part\_Func($\omega^i$, $a^i$)
	
	$Z$ initialization
	
	\For{$i$ $\mathrm{in}$ $V \setminus \{\left[(L-1)\times L +1\right] \bigcup L^2\}$ }{
		$Z_{i\left[(L-1)\times L +1\right]}=\mathrm{Calc\_Z_{ij}}(i,(L-1)\times L +1,a,\omega)$
		
		$Z_{iL^2}=\mathrm{Calc\_Z_{ij}}(i,L^2,a,\omega)$
	}
	
	$Z_{\left[(L-1)\times L +1\right]L^2}=\mathrm{Calc\_Z_{ij}}((L-1)\times L +1,L^2,a,\omega)$
	\\ 
	\tcc*{ and $\mathrm{Calc\_Z_{ij}}(i,j,a,\omega)$ is described next} \hspace{5mm}
	
	\textbf{procedure} $\mathrm{Calc\_Z_{ij}}(i,j,a,\omega)$
	
	\eIf{$Z_{ij} \textnormal{ is already in } Z$}{
	return $Z_{ij}$}{
	\eIf{$\mathrm{city\_block\_distance}(i,j)=1$}{
		$Z_{ij}=1$
		
		$Z_{ji}=1$
		
		return $Z_{ij}$}{
		\textit{up}, \textit{down}, \textit{left}, \textit{right}=$\mathrm{calculate\_direction}(i,j)$ 
		\tcc*{Each value is \textbf{true} if $i$ is in that direction respect $j$}
		\begin{align}
 			Z_{ij} & = \textit{up} \times \textnormal{e}^{-\beta(a_{j_{\uparrow}}+\omega_{jj_{\uparrow}}) }\times \mathrm{Calc\_Z_{ij}}(i,j_{\uparrow},a,\omega) \nonumber \\
			&+ \textit{down} \times \textnormal{e}^{-\beta(a_{j_{\downarrow}}+\omega_{jj_{\downarrow}}) }\times \mathrm{Calc\_Z_{ij}}(i,j_{\downarrow},a,\omega) \nonumber \\
			&+ \textit{left} \times \textnormal{e}^{-\beta(a_{j_{\leftarrow}}+\omega_{jj_{\leftarrow}}) }\times \mathrm{Calc\_Z_{ij}}(i,j_{\leftarrow},a,\omega) \nonumber \\
			&+ \textit{right} \times \textnormal{e}^{-\beta(a_{j_{\rightarrow}}+\omega_{jj_{\rightarrow}}) }\times \mathrm{Calc\_Z_{ij}}(i,j_{\rightarrow},a,\omega) \nonumber
		\end{align}
		
		$Z_{ji} = Z_{ij}$
		
		return $Z_{ij}$
}}	
\tcc{Notice that $Z$ is a data structure and $Z_ij$ are values stored inside $Z$ using the adequate indexes}
\end{algorithm}

This approach is more efficient that obtaining every partition function following equation \eqref{eq:Z}. Instead of adding over every possible path the already calculated partition functions are used to avoid redundant computations. This reduction in calculations becomes more significant as city side grows. For instance, if a partition function is calculated for points distant $q=K$ steps horizontally and $p=K$ steps vertically, the brute force computation results in $\sim 2^{2K}$ from an Stirling's approximation of \eqref{eq:npaths}. Exploiting the our recursive procedure this results in $\sim K^2$ operations.




\nohyphens{The final step in the gradient descent is the updating of the weights in the direction of the gradient in order to maximize the log likelihood. This can be done with a fixed learning rate like the stochastic learning rate or with an adaptative step, like Adam gradient descent algorithm or similar \cite{ruder_overview_2017}. The gradient steps are carried up to when a given tolerance is achieved or a maximum number of loops are reached. Typically the tolerance is fixed as a small value for the norm of the gradient.}


\section{Results}
\label{sec:results}

We now show some experiments to test the precision of our inference methodology with synthetic data from an artificial city. We will consider a grid-like city of $N=L\times L$ nodes ($L=12$) and a link weights extracted from a scaled Gaussian distribution 
\[ w_{i,j} = \beta \omega^0_{i,j} \quad \text{where } \omega^0_{i,j} \sim N(\mu=0,\sigma^2= 1 ).\]
We do not consider node weights, since they are equivalent to a gauge transformation of the links.


In artificial city is possible to take several paths between each origin and destination. 
The probability of each paths can be computed using \eqref{eq:PP} given a set of weights.
From this artificial city we generate a data set of triplets $\D$. 
Each triplets is composed by an origin, an intermediate point and destination.
The length of the set, $|\D|$, corresponds with the total of triplets.
This set of triplets are similar to the data set of sparse trajectories of mobile phone registers after be split.

\begin{wrapfigure}{r}{0.5\textwidth} 
	\includegraphics[scale=0.55]{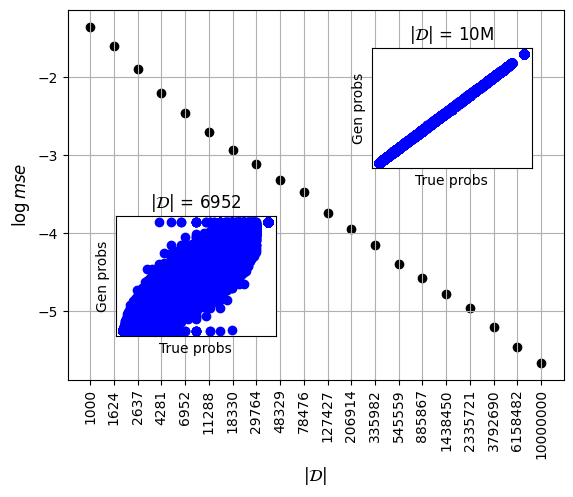}
	\caption{\label{fig:len_mse} Log mean square error for several length of $\D$, in an homogeneous city, $\beta=0$. True probabilities vs model generated probabilities for particular values of $\D$ in inner plots}
\vspace{-1cm}
\end{wrapfigure}

At this point, we have a set of paths for each origin destination pair and the probability to take each path. The probability to use $k$ in a $i-j$ path is easily obtainable summing over every $i-j$ path that includes $k$. Then, we can pick origin-destination pairs and sample the data according the distribution created. 

There are two main parameters to consider in order to perform a proper evaluation.
The first is $|\D|$, the length of the set. If our procedure is correct, the quality of our inference should grow towards perfection when the number of samples in $\D$ is large.
The second parameter is $\beta$, this value represents the inverse of the temperature, taking a direct effect over the shape of the distribution of the paths probabilities. For $\beta=0$ (infinite temperature) every path has equal probability with independence of the values of $\omega$ and $a$. With an increase of $\beta$ (temperature decrease) is expected that the path with the lower Hamiltonian \eqref{eq:ham} became more probable respect the others with same origin and destination.

Firstly, we explore the influence of $|\D|$ over mean square error of the probabilities generated with the inferred weights. For that, we set $\beta=0$ and perform the procedure presented in the previous section to obtain a set of weights from a random ones.

Once the new weights are inferred we comput:
\begin{equation}
	mse=\frac{1}{N}\sum_{(ikj) \in \D}\sqrt{(p_{r_{(ikj)}})^2-(p_{i_{(ikj)}})^2}
\end{equation}

\noindent where $N$ is the total of $ikj$ in the graphs and $p_{r_{(ikj)}}$ and $p_{r_{(ikj)}}$ stands for the real and inferred probability for the trio $ikj$, respectively.

Figure \ref{fig:len_mse} shows the results of the logarithm of mean square error with the grow of $|\D|$. As excepted the error decreases with an increases of length $\D$. 

\begin{figure}[!htb]
\begin{minipage}{0.48\textwidth}
	\includegraphics[scale=0.6]{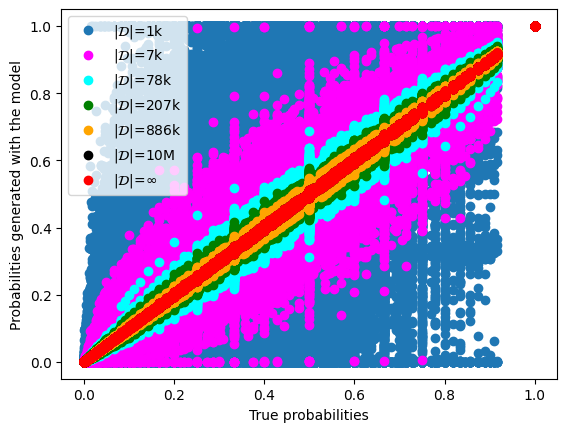}
	\caption{\label{fig:pr_pi} True probabilities against probabilities generated using model’s parameters}
\end{minipage}
\begin{minipage}{0.45\textwidth}
	\includegraphics[scale=0.5]{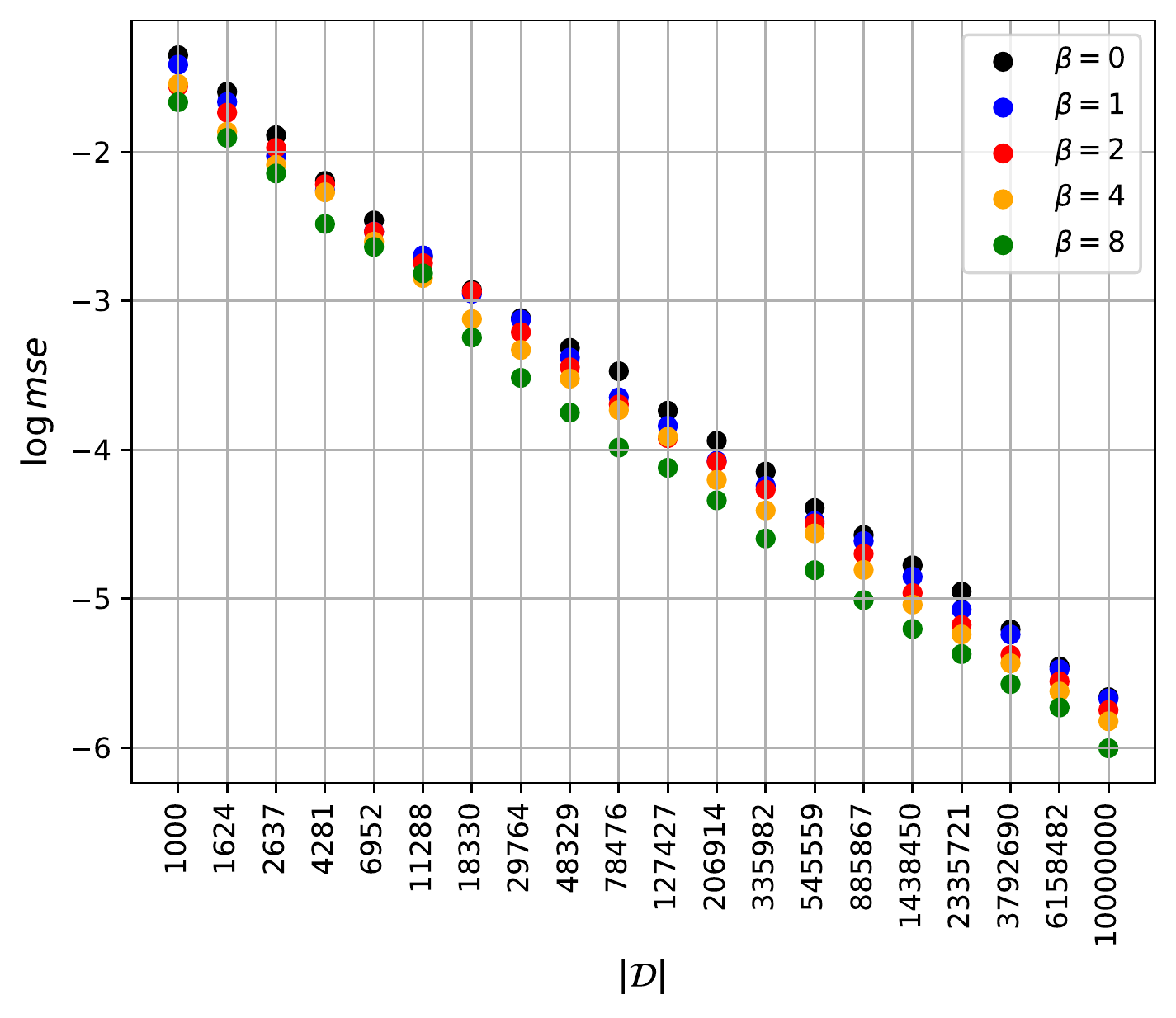}
	\caption{\label{fig:len_mse_beta} Mean square error varying length of $\D$, using different $\beta$}
\end{minipage}
\end{figure} 

In the inner plots of figure \ref{fig:len_mse} is shown the behavior of $p_{r_{(ikj)}}$ against $p_{r_{(ikj)}}$ for every $ikj$. Notice that with a bigger number of trios the distribution of dots became similar to the curve $x=y$, as expected. This is shown with more detail in figure \ref{fig:pr_pi}, when also is considered the case with length of $\D$ infinite. Notice, that to assume this case is needed to set $n$ and $m$ with the value of its convergence with the increase of $|\D|$.

\nohyphens{The analysis done in figure \ref{fig:len_mse} can be extended to others values of $\beta$ to prove that the same behavior is obtained in a non-homogeneous city. This is shown in figure \ref{fig:len_mse_beta}, where using the same set of discrete $|\D|$ the values of the logarithm of mse are plotted from different values of $\beta$.
Particularly, $\beta=0$, $\beta=1$, $\beta=2$, $\beta=4$ and $\beta=8$.}

%

%

\section{Conclusions}
\label{sec:conclusions}

We introduce the City Path Tomography problem and show that it can be solved in a very simplified toy model of phone users moving in a grid-like city. The likelihood of the observed data is maximized in an efficient way, thanks, mostly to the simplicity of the city and the assumption of users moving along non returning paths. As expected, the inference is more accurate as the amount of data growths. 

The success of this methodology act as a proof of concept. It is a first step to attempt the more challenging situation of realistic cities and travelers. The challenge in that case is two fold: first describe a model that is consistent with real human behavior in a city, and second, solve it (probably resorting to approximate stat mech methods). 

\section{Acknowledgments}
\nohyphens{The research presented in this publication received funds from the Office of International Funds and Projects Management under the code PN223LH006-007.}

\bibliographystyle{unsrt}
\bibliography{references}

\end{document}